\documentclass[10pt]{article}
\renewcommand{\baselinestretch} {1.5}
\title{\textbf{New Indications of Changing the Regime of Multiple Production at
 Superhigh Energies}}
\author{B.N. Kalinkin$^{1}$ and
Yu.F. Gagarin$^{2}\thanks{e-mail: y.gagarin@pop.ioffe.rssi.ru}$}

\date {}

\begin{document}
\vspace{2cm}
\maketitle

\noindent$^{1}$ Joint Institute for Nuclear Research, 141980 Dubna,
Moscow Region, Russia\\
\noindent$^{2}$
Ioffe Physico-Technical Institute of the Russian Academy of Sciences,
194021 St.-Petersburg, Russia\\

\noindent PACS REf: 96.40

\begin{abstract}
\noindent The effect of alignment of secondary particles (products of
hadron interactions at superhigh energies
$E^\mathrm{{in}}\geq5\cdot10^{6}$ GeV) as a "fan", which has been
observed in cosmic rays, is  analysed.  It is shown that its main
specific features are well described within the model that we proposed
twenty years ago  to explain sharp anomalies in the process of multiple
production, which are observed in the same energy range. This model
assumes that quarks have internal massive degrees of freedom. Some
consequences of the used approach are discussed, which may be important
for the multiple process in the above-mentioned region.  \end{abstract}
\section{Introduction}
In processes at superhigh energies ($E^\mathrm{{in}}>10^{6}-10^{7}$
GeV), earlier in cosmic rays [1] and now at new accelerators [2], there
have been observed serious deviations from the particle production
regime.  First of all, it concerns a multiple process, which was
observed at "accelerator" energies up to SPS energy and could
successfully been described, for instance, in the framework of a
phenomenological model of gluon dominance (MGD) for pp-, pA- and AA-
collisions [3,4] or in its detailed version, a model of hadron strings
(MHS) (see, in particular, [5]) as well as in the framework of a
standard QCD approach.

These deviations imply a sharp increase in multiplicity,
average transverse momentum $<p_{\bot}>$ of produced particles and a
change of qualitative composition of secondary hadrons in favour of
more massive ones [1]. And this is not the end of the chain of this
kind of facts.

Recently [6], in studying the interactions of cosmic
rays in stratosphere at "Concord" an event has been observed in which
about 200 vigorous $\gamma$-quanta ($E_{\gamma}\geq200$ GeV) with
very small deviations turned out to be in the same plane, forming a
"fan" (the term is from [6]).

However, one cannot accept this
observation as an exceptional or exotic one since similar events have
already been observed in the experiment "Pamir" when studying $\gamma$-hadron
superfamilies of particles generated by the primary hadron with energy
around $10^{7}$ GeV and higher [7]. First important conclusions have
been made in [7].

First, the effect of alignment of particles as a "fan"
has a threshold nature and arises at primary energies exceeding
$10^{7}$ GeV.

Second, at these energies a portion of events
demonstrating the alignment of particles is not small and amounts to
20--40\% of all inelastic collisions.

Third, in the process leading to
the alignment, the central role should be played by heavy particles
with mass of an order of hundreds GeV.  Otherwise, as a result of
cascading in the atmosphere the picture of alignment should be smeared.

Finally, attempts to explain the effect using traditional schemes based
on the MHS representation of breaking of hadron strings which arise in
the quark-quark interaction, did not give any positive results (for
discussion see [7]).

Consequently, alongside with the earlier
established sharp anomalies in the behaviour of secondary particles,
which arise at superhigh energies [1,2], we face one more anomaly,
qualitative change of the shape of their angular distribution.

\section{Alignment effect - a new evidence for the qualitative
change of regime of multiple production.}

The problem arisen can be overcome either by making the above-mentioned
approaches more complicated supplementing them with sophisticated superstructures
or looking for a qualitatively new solution. Both the ways have right to exist.
We choose the second way and give interpretation of events of the type of a
"fan", which is based on the model of a new regime for an inelastic interaction
of nucleons at superhigh energies, which we suggested 20 years ago [8].
According to it , violation of the standard MGD behaviour of characteristics
of a multiple production process is due to the fact that at high energies in
collisions close to head-on ones internal degrees of freedom of quarks are
"defrosted", i.e., their internal structure manifests itself. This results
in absorption of a great amount of collision energy by internal massive degrees
of freedom, their excitation. As a result, there may be a situation when an
ordinary "leading"effect, in complete correspondence with experiment [1,9],
disappears and almost all the energy per valence quarks in colliding hadrons
turns into excitation energy of a hadron object produced, on average resting
in the center of mass system. This object was then called the
Q-cluster. The proposed model allows one to explain (for details see
[8]) basic of the observed anomalies in the behaviour of multiple
production characteristics [1,2,9].

One can easily be convinced that
the appearance of the effect of alignment of particles as a "fan" at
superhigh energies can easily be explained in the framework of the
model [8].

Indeed, if as a result of collision there appears a
Q-cluster as a unified hadron system, from the conservation laws
 there automatically follows that besides excitation energy it must
 have an angular momentum as well. The fact that the unified hadron
 system may in principle possess an angular momentum was established by
 us almost 10 years ago while studying angular distribution of
 particles produced in cumulative reactions at smaller energies [10].

Let us estimate the angular momentum for the case of a "fan" effect.
It is obvious that a unified system can be formed if the impact
parameter $"b"$ in a collision of nucleons does not exceed the value

\begin{equation}b^\mathrm{{max}} \approx l \approx0.5\mbox{ fm
},\end{equation}

\noindent where $l$ is the value of smearing the boundary of
distribution of quarks in a nucleon, as in a quark bag. (This issue has
in detail been considered in [10], and in [11] - relation (28)).
Averaging over impact parameters in the interval
$0\leq{b}\leq{b^\mathrm{{max}}}$ gives for the average value $<b>$:

\begin{equation}<b>\approx\frac{2}{3}b^\mathrm{{max}}\approx0.33\mbox{
fm }.\end{equation}

\noindent For the collision energy we take the value from [6], i.e.,

\begin{equation}E_\mathrm{{p}}^\mathrm{{in}}=10^{7}\mbox{ GeV
}.\end{equation}

\noindent Then, an approximately average value of the angular momentum
of the Q-cluster $<\mathrm{L}_\mathrm{{Q}}>$  should be

\begin{equation}<\mathrm{L}_\mathrm{{Q}}>\approx\frac{<b>\sqrt{s}}{2\hbar}\approx3\cdot10^{3}.\end{equation}

\noindent The allowance for energy carried away by the gluon fields of
colliding nucleons would lead to a coefficient of an order of 0.7. This
correction does not considerably change the result. Moreover, it is not
obvious whether this correction is necessary or not.

A huge value of $<\mathrm{L}_\mathrm{{Q}}>$ in (4) would certainly lead
to a quasiclassical situation - particles and generators of jets
leaving the Q-cluster should be arranged in the R plane normal to the
angular momentum vector $\vec{\mathrm{L}}_\mathrm{{Q}}$ and going
through the collision axis.  The situation in c.m.s. is shown in Fig.1.
It is obvious that in the laboratory system this scheme of particle and
jet dispersion qualitatively corresponds to the observed effect of a
"fan" (especially if one takes into account that products with large
energy, more than 200 GeV, are registered).

To verify a high degree of coplanarity of dispersion near the R plane,
one has to show that the value of smearing the angular momentum
distribution near $\theta=\pi/2$ is rather small due to the
presence of quantum $"q"$ corrections. It is known that the halfwidth
$\Delta\theta$ of distribution over $\theta$
corresponding to
the square of module of the wave function of the rotation motion
$\mid\Theta_{ll}\mid^2$ near $\theta=\pi/2$ in the quasiclassical
approximation is equal to

\begin{equation}\Delta\theta_{q}\approx\sqrt{\frac{1}{2l}}.\end{equation}

If in the considered event one observes, for instance, 3 or 5 jets, each
of them on average carries away from the Q-cluster the angular
momentum $\mathrm{L}_\mathrm{{Q}}/3$ or $\mathrm{L}_\mathrm{{Q}}/5$,
respectively.  Then, on the basis of (4) and (5) we get for the
smearing $\Delta\theta$

\begin{equation}\Delta\theta_{q}\approx\sqrt\frac{1}{\frac{2}{3}\mathrm{L}_\mathrm{{Q}}}
\div\sqrt\frac{1}{\frac{2}{5}\mathrm{L}_\mathrm{{Q}}}.\end{equation}

\noindent Even for ten-jet events this smearing amounts to $1.7^{o}$.

\noindent The estimate (6) quantitavely confirms a high degree of
coplanarity of jet and particle dispersion, according to the picture of
the Q-cluster decay suggested by us in [8].  Consequently, the image of
a "fan" in this sort of events is justified.

\section{Interpretation of basic features of fan-type \\events within
our model}

\noindent So, in the framework of our model [8] the effect of alignment
has a natural and quite transparent interpretation. Using it we can
comment on the results [7] of experimental study of the effect in the
order they are presented in paragraph \textbf{1}.

First, the threshold nature of the effect automatically follows from
the very essence of the model [8] as it is assumed that at sufficiently
high energies internal degrees of quarks of colliding nucleons are
"defrosted" (see paragraph \textbf{2}).

Second, the adopted sceme gives a
considerable value for a portion of events with alignment to a total
number of inelastic collisions
$\delta=\sigma_\mathrm{{fan}}/\sigma^\mathrm{{inel}}$.  The lower
estimate $\delta^\mathrm{{min}}$ can be obtained if for calculation of
$\sigma_\mathrm{{fan}}$ we again use the quasiclassical approximation

\begin{equation}\sigma_\mathrm{{fan}}^\mathrm{{min}}\approx\pi(b^\mathrm{{max}})^2
\approx8\mbox{ mb }.\end{equation}

Estimate (7) assumes that "transparency" (in terms of nuclear optics) is
close to zero in this sort of collisions. This assumption is justified as,
according to our model,  collisions  of  nucleons  with  the  impact
parameters $"b"$ in the interval
$0\leq{b}\leq{b^\mathrm{{max}}}$  are relevant to destruction of all
three pairs of quarks and disappearance of the leading effect. Assuming
that $\sigma^\mathrm{{inel}}\approx40$ mb at $E^\mathrm{{in}}=10^{7}$
GeV (according to the data given in [1]) the increase in
$\sigma^\mathrm{{inel}}$ in comparison with its value in the
"accelerator" region may amount to 30\% get $\delta^\mathrm{{min}}$ to
be equal to

\begin{equation}\delta^\mathrm{{min}}\approx0.2.\end{equation}

It  is obvious that the contribution to the effect may also come from
collisions accompanied by destruction of two pairs of quarks, which correspond
to somewhat larger values in $b^\mathrm{{max}}$ and larger
"transparency".  Therefore, a real value of the portion of $\delta$ may
turn out to be 1.5-2 times as large as the value (8). Consequently,
there is no contradiction with experiment here as well.

Third, in our earlier
 paper [8], on the then available data [1] on the value of the
 threshold energy $E^\mathrm{{in}}_\mathrm{{th}}$ corresponding to
transition to a new regime, and on the assumption that quark
destruction can simply be represented as production of a pair of
 intermediate unstable particles (let us call them "H"-particles -
 heavy particles), we derived for their mass
 $\mathrm{M_{H}}=\sqrt{s_{qq}}/2\approx50$ GeV.  Consequently, the
model [8] qualitatively (and automatically like above) agrees with the
conclusions of [7].  However, at present a numerical estimate of mass
 $\mathrm{M_{H}}$ has to be revised as in the recent [9,12] modern
energy values are given starting from which a new regime acquires a
stable character, namely
$E^\mathrm{{in}}\geq5\cdot10^{6} - 2\cdot10^{7}$ GeV.
This value much
larger than the one we used earlier [8]. Making calculation
of [8] with a new value for $E^\mathrm{{in}}_\mathrm{{th}}$ we obtain

\begin{equation}\mathrm{M_{H}}\approx400\mbox{ GeV }.\end{equation}

\noindent The author of [7], following other reasonings, considers such
a value for $\mathrm{M_{H}}$  to be preferable.

\subsection{}

Having the value for the mass of H-particles
$\mathrm{M_{H}}\approx400$ GeV and for the Q-cluster
 temperature $\mathrm{T_{Q}}\approx3$ GeV, which has been obtained
earlier in [8], one can roughly estimate a mean value of the angle
$\Delta\theta_\mathrm{{T}}$ of deflection of H-particles from the plane
R (see Fig.1) in the laboratory system of coordinates

\begin{equation}\Delta\theta_\mathrm{{T}}\approx(\arctan\frac{<p^\mathrm{{H}}_{\bot}>}
{<p^\mathrm{{H}}_{\parallel}>})_\mathrm{{lab}}.\end{equation}

Let us first note that kinetic energy of an H-particle in the rest frame
of the excited Q-cluster in thermodynamical approximation is of an
order of $E_\mathrm{{kin}}^\mathrm{{H}}\approx3\mathrm{T}/2\approx4.5$
GeV.  Hence, for the velocity $v_\mathrm{{H}}$ in the same system we
get

\begin{equation}\frac{v_\mathrm{{H}}}{c}\approx0.15.\end{equation}
Consequently, H-particles in the Q-system are extremely
nonrelativistic.
Taking into account that $\mathrm{M_{H}}\gg{\mathrm{T_{Q}}}$ one can
easily obtain for $<p^\mathrm{{H}}_{\bot}>$:

\begin{equation}<p^\mathrm{{H}}_{\bot}>\approx\sqrt{\frac{\pi
\mathrm{M_{H}}\mathrm{T_{Q}}}{2}}.\end{equation}
For $<p^\mathrm{{H}}_{\parallel}>$  (neglecting a small velocity
$v_\mathrm{{H}}$ in the Q-system) we have

\begin{equation}<p^\mathrm{{H}}_{\parallel}>_\mathrm{{lab}}\approx
\mathrm{M_{H}}\cdot\gamma_\mathrm{{Q}},\end{equation}
where $\gamma_\mathrm{{Q}}$ is the Lorentz factor of the Q-cluster in
the lab. system

\begin{equation}\gamma_\mathrm{{Q}}\approx1.2\cdot10^{3}.\end{equation}
Since $<p^\mathrm{{H}}_{\parallel}>\gg<p^\mathrm{{H}}_{\bot}>$

\begin{equation}\Delta\theta_\mathrm{{T}}\approx\frac{<p^\mathrm{{H}}_{\bot}>}
{<p^\mathrm{{H}}_{\parallel}>}\approx9\cdot10^{-5}\approx15''.\end{equation}
This value exceeds the quantum smearing
$(\Delta\theta_{q})_\mathrm{{lab}}\approx(\Delta\theta_{q})_\mathrm{{cms}}/\gamma_\mathrm{{Q}}\approx1''- 5''$
(see paragraph \textbf{2}).

For comparison with experiment let us use the data [6] on the dimension and mutual arrangement of halos from
the groups of $\gamma$-quanta (see the figure therein) as well as the
assumed distance ($\approx100$ m) from the primary act to a registering
set up. The estimate $\Delta\theta$ from these data gives

\begin{equation}\Delta\theta_\mathrm{{exp}}\approx4''- 6''.\end{equation}
Indeed, for a realistic comparison of this value $\Delta\theta$ with
the model estimate $\Delta\theta_\mathrm{{T}}$ in eq.(15) one should
take into account obvious facts.

First, limited accuracy of experimental determination of the distance between the point of a primary
interaction (appearance of the Q-cluster) and a registering
equipment.  Moreover, selection of secondary registered particles by
the criterion of a large value of their summed energy certainly
decreases an average angle of particle dispersion near the plane of a
"fan".

Second, it is obvious that the model estimate of the quantity
$\Delta\theta_\mathrm{{T}}$  is rather rough.  It is most probably an
estimate from above as the allowance for a large value of the rest-mass
of H-particles would lead to a noticeable decrease in the temperature
value.

Therefore, one should expect agreement of $\Delta\theta_\mathrm{{exp}}$
with $\Delta\theta_\mathrm{{T}}$, in the best case, only in the order
of magnitude.  In this sense the result obtained can be considered
satisfactory.

\subsection{}

Important conclusions can be drawn from comparison of
the results of "Pamir" experiments with those obtained on the
"Concord".  They correspond to distances $h_\mathrm{{Q}}$ between the
point of primary interaction (by assumption, the point of production of
a Q-cluster) and a recorder which are highly different in magnitude;
for the "Pamir" $h_\mathrm{{Q}}\approx3$ km and for the "Concord"
$h_\mathrm{{Q}}\approx100$ m.

We note that events recorded at the "Pamir" are characterized by three
halo-spots from beams of $\gamma$-quanta.  The event registered at the
"Concord" consists of a double amount of aligned beams. Qualitatively,
this ratio may happen to be not accidental. The chain length of
$\gamma$-beams  observed in the experiment "Concord" approximately
equals 41 mm. If this "fan" was observed at a distance of 3 km from
the point of primary interaction (i.e., in the conditions of the
"Pamir"), the length of that chain would be 123 cm.  This value is
comparable with horizontal sizes of the "Pamir" recorder.
Consequently, the probability for the "Pamir" recorder to fix only
separate fragments of the type "fan" discovered on the "Concord" is
sufficiently large.

It is instructive to examine the situation in reverse order. The length of
the chain of three halos in the "Pamir" experiment (at least, of those cited
in [7] equals $1.5 - 2.0$ cm. If
these halos were registered at $h_\mathrm{{Q}}\approx100$ m (i.e. at the
"Concord") rather than at $h_\mathrm{{Q}}\approx3$ km, all of them
would be in an interval not larger than 0.7 mm. However, the accuracy
of data reported in [6] does not exceed 1 mm. Therefore, the assumption
that 5 beams of $\gamma$-quanta in the "Concord" event are produced not
by 5 but a much larger number of H-particles (perhaps, 2 -- 3 times) can
be taken as realistic.  Once the process of decay of initiators is
completed, the picture with a large number (up to 10 -- 15) of hadron
jets can be expected.

Since the events under consideration remain aligned up to distances $
h_\mathrm{{Q}}=3$ km, it can be assumed that the lifetime of
H-particles is of the order

\begin{equation}\tau_\mathrm{{o}}^\mathrm{{H}}\approx\frac{3\mbox{ km
}}{c\cdot\gamma_\mathrm{{Q}}}\approx0.8\cdot10^{-8}\mbox{ sec }
\end{equation}

\noindent and, therefore, the mean free path of H-particles
before decay, $<l_\mathrm{{H}}>$, in experiments on colliding $pp$ and
$p\bar{p}$ can be estimated as follows:

\begin{equation}<l_\mathrm{{H}}>\geq\tau_\mathrm{{o}}^\mathrm{{H}}\cdot
v_\mathrm{{H}}\approx40\mbox{ cm}.\end{equation}

As for the energetic $\gamma$-quanta of observed halos, they appear, as
we think, from the energetic part (in accordance with the experimental
conditions) of bremsstrahlung of $\pi^\mathrm{{o}}$,
$\rho^\mathrm{{o}}$ and other mesons by H-particles with their
subsequent decay containing the mode with $\gamma$-quanta.

\subsection{}

There is a related problem of the appearance of
"knee" in the spectrum of the photon-electron component observed
experimentally at the same energies.  The authors of ref. [9] have
assumed that this knee is due to
\noindent vanishing leading hadrons.  In our opinion, this
assumption is necessary but not sufficient.  Indeed, when multiple
processes go over to a new regime, the part of energy spent onto
production of new particles (the "thermal" energy $W_\mathrm{{T}}$)
strongly increases, which can be seen in Fig.2. Curve 1 describes the
behaviour of $W_\mathrm{{T}}$ fitted to the experiment up to
$E^\mathrm{{in}} = (2 - 3)\cdot10^{5}$ GeV (see reviews in [4]) and
extrapolated onto the region of $E^\mathrm{{in}}>10^{6} - 10^{7}$
GeV.  Curve 2 is the dependence $W_\mathrm{{T}}(E^\mathrm{{in}}$)
described by the model [8].  Curve 3 describes this dependence with
the inclusion of the mechanism of breakdown of quarks, which
demonstrates a strong growth of the energy release into the channel of
multiple production at initial energies $E^\mathrm{{in}} > 10^{6}$ GeV.
Therefore, to coordinate the appearance of the knee in the
$e,\gamma$-spectrum with the disappearance of leading hadrons, it is to
be assumed that the energy losses of new H-particles (up to their
decay), which take a huge part of the energy $W_\mathrm{{T}}$, are much
lower than those of usual primary baryons.  This assumption is
justified since H-particles being small in sizes and huge in mass can,
only with a small cross section, transfer a small part of their energy
to baryons in an atmospheric gaseous mixture.

It is possible that a quark consists of three superheavy fermions,
H-particles, in analogy with the three-quark structure of a
conventional baryon, nucleon. This idea is rather attractive,
though speculative.

So, within the model [8], the
assumption that the knee in the $e,\gamma$-spectrum appears due to
disappearing leading hadrons becomes not only necessary but sufficient,
as well.

\section{Some consequences of the adopted picture}

We complete the exposition of results obtained in [8] and in this paper with
some important consequences from the model [8].

\subsection{}

It was noted in [1] that the cross section of
inelastic interaction grows when the new regime sets in. This means in
terms of the hadron optics that the "transparency" of a nucleon-target
decreases for an incident nucleon, but just this is , in accordance
with the model [8], a consequence of "defreezing" of new degrees of
freedom of quarks and formation of a Q-cluster followed by the
disappearance of "leading" effect. Therefore, it is not surprising that
a relative growth of $\sigma^\mathrm{{inel}}$ at energies under
consideration turns out to be close in magnitude to the part of aligned
events $\sigma_\mathrm{{fan}}$.

\subsection{}

Approximately, it is not difficult to estimate the
size of transition to a new regime in the energy scale
$E^\mathrm{{in}}$.  In the c.m.s. a pair of quarks in colliding
nucleons (start of a new regime $E^\mathrm{{in}}_\mathrm{{th}}$) to be
broken requires an energy 3 times as low as that for breaking all 3
pairs of quarks and formation of a complete Q-cluster (full inclusion
of a new regime $E^\mathrm{{in}}_\mathrm{{Q}}$). Therefore it is clear
that the domain of transition in the lab. system should approximately
cover an one-order interval, i.e.
$\log(E^\mathrm{{in}}_\mathrm{{Q}}/E^\mathrm{{in}}_\mathrm{{th}})\approx 1$.  Boundaries of that domain should be slightly smoothed due to
one-quark internal motion inside nucleons before collision, collective
motion of the whole quark system with respect to the total gluon field,
and due to possible excitation of quarks not followed by their decay.

This result of the
model [8] is also in agreement with approximate data given in [1] on
the size of transition to a new regime on the
$E^\mathrm{{in}}_\mathrm{{lab}}$ scale.  If modern data on a larger
value of the initial energy, compared to its previous estimates were
taken into consideration, the whole picture would only shift towards
higher initial energies (qualitatively, it is shown in Fig.2).

\subsection{}
Since an effective value of  the rotational energy of
a Q-cluster can be a half of its total internal energy, it can be
assumed that the estimate of $\mathrm{M_{H}}$ given by (9) is somewhat
large and its realistic value can be half of that value. Therefore we
expect $\mathrm{M_{H}}$  to be in the interval

\begin{equation}200\mbox{ GeV }\leq \mathrm{M_{H}}\leq
400\mbox{ GeV }.\end{equation}
At present, it is difficult to make a more accurate estimation,
however,  a possible decrease of  the estimate (19) will not
essentially influence our results.

It is of interest to point to the following fact: in recent experiments on
colliding beams of electrons and protons at HERA [13]
an anomaly has been discovered in $ep$-scattering which testifies, in
experimenters'opinion, to the structure of a proton of order of 0.001
of its size, i.e. $\lambda_\mathrm{{M}}\approx10^{-3}<r_{p}>$. This
observation does not contradict the assumption we have made in [8]
concerning degrees of freedom of a nucleon deeper than quarks. The mass
of the supposed structure is estimated to be

\begin{equation}\mathrm{M}\geq\frac{\hbar}{\lambda_\mathrm{{M}}}\approx
250\mbox{ GeV }.\end{equation}
We see that estimates (19) and (20) obtained on
the basis of qualitatively different processes provide close values for
$\mathrm{M_{H}}$. This comparison is sure to be tentative since the
experiment [13] is not yet completed, and accidental elements are not
excluded.

\subsection{}

As a result, from our analysis we can make preliminary
and speculative conclusions on the properties of H-particles:

\begin{itemize}
\item
     their mass is in the interval (19);
\item
     the lifetime is of order
$\tau_\mathrm{{o}}^\mathrm{{H}}\approx10^{-8}\mbox{ sec};$
\item
      H-particles are most likely fermions (heavy bosons with masses in
      the interval 610 GeV/c$^{2} > m > m_{W}$ are not yet
      discovered [14]);
\item
      H-particles are strongly interacting
      particles since the construction of a quark from them requires a
      huge mass defect (of an order of $ 400 - 800\mbox{ GeV })$ and
      this is necessary for realization of a Q-cluster as an
      intermediate highly excited system that exists for a
      certain time (otherwise, the event with alignment becomes
      a mystery).
\end{itemize}

From the above properties of
H-particles we may assume that they can be unstable superheavy baryons.

\section{Conclusion}

A number of sharp anomalies in the behaviour of characteristics of multiple
production discovered in experiments with cosmic rays at ultrahigh energies
and a striking effect of the "fan" with the highest degree of complanarity can
be described within a unique mechanism formulated in [8,15].

The proposed scheme can be realized via the basic assumption made in [8] that
quarks have internal structure constructed on massive degrees of freedom and
in strong excitation they can destroy decaying into heavy unstable particles
with a mass of hundreds of GeV. From a gnosiological standpoint, the
assumption that quarks are not the latest indivisible elementary bricks
in the structure of matter is not crazy, which follows from the whole
history of physics.

Accuracy of the experiments under discussion and a
qualitative estimate of the initial energy is not high, therefore, they
should separately be estimated thoroughly. However, they together, at
least, qualitatively can be given a common description and this
testifies to the validity of the model proposed in [8]. In any case
analysis of new data in [16] is not in conflict with our conclusions.

We hope that
considerations we made in [8] and in this paper concerning the results
available on the study of processes at ultrahigh energies will get
further development; we mean progress in the experimental technique
with cosmic rays  (satellites of the Earth and stratospheric airplanes)
and start-up of new accelerators (for example, LHG). Figure 2 illustrates a rapid decrease
of the gap between energies attained at modern accelerators and the
energy region of sharp anomalies observed in experiments with cosmic
rays. As can be seen from Fig.2, energies at the FNAL tevatron in the
scale of $E_\mathrm{{p}}^\mathrm{{in}}$ approach the threshold of new
regime.  In our opinion, a direct evidence for this statement comes
from observed considerable deviations of the dependence of the yields
of jets, $J/\Psi$- and $\Upsilon^\mathrm{{o}}$-particles on $p_{\bot}$
at $\sqrt{s_{p\bar{p}}}=1.8\mbox{ TeV }$ [2] from the expected one.
Spectra get much more hard, which in terms of thermodynamics
corresponds to the start of anomalous growth of the temperature of an
intermediate system.  In the framework of the discussed picture, this
situation is natural:  pumping of the energy into an intermediate
system begins fast growing as a result of defreezing the internal
degrees of freedom of quarks leading first to their excitation and then
to disintegration (earlier in [8], a behaviour like that was
demonstrated for the distribution of extensive atmospheric shower
"stems" over $p_{\bot}$).

Further
experimental verification of the conception discussed in [8] and in
this paper will support the idea of a quark as a structural object
functionally analogous to a quasiparticle, an object widely used in
different fields of physics.  This perspective seems still more
realistic in view of recent important results: for the first time, in
semiconductors discovered are quasiparticles with fractional charges
equal to 1/3 of the standard charge [17].

As we have mentioned in [8], the idea of quarks being quasiparticles can
drastically change the view
of the essence of confinement (quarks, quasiparticles cannot exist
beyond a system, a hadron, whose local properties they express) and can
impose the limits on the use of the asymptotic-freedom approximation
(evidently, it is to be employed only in the subthreshold region of
energies, i.e. at $\sqrt{s_{p\bar{p}}} < 1-2 \mbox{ TeV }$).

\noindent\textbf{Acknowledgements}

\noindent The authors are gratefull to prof. V.A. Meshcheryakov for discussions.

\noindent\textbf{Figure captions}

\noindent\textit{Fig. 1.} The scheme of emission of decay products of a
$Q$-cluster with a large angular momentum in the c.m.s.

\noindent\textit{Fig. 2.} The $E_\mathrm{{p}}^\mathrm{{in}}$ - dependence of the energy
spent for production of new particles in a $pp$-collision. Arrows
denote values of $E_\mathrm{{p}}^\mathrm{{in}}$ equivalent in energies
$\sqrt{s}$ reached at colliding beams of accelerators ISR, SPS and the
FNAL tevatron: 60 GeV, 540 GeV and 1.8 TeV. Curve 1 -
the $W_\mathrm{{T}}(E^\mathrm{{in}})$ dependence according MGD
[4] up to $(2-3)\cdot10^{5}$ GeV and extropolation onto the region
$E^\mathrm{{in}} > 10^{6} - 10^{7}$ GeV; Curve 2 -
the $W_\mathrm{{T}}(E^\mathrm{{in}})$ dependence according
Q-cluster model [8]; Curve 3 - the $W_\mathrm{{T}}(E^\mathrm{{in}})$
dependense with the inclusion of the mechanism of breakdown of quarks.
\pagebreak
\renewcommand\baselinestretch{1.428}
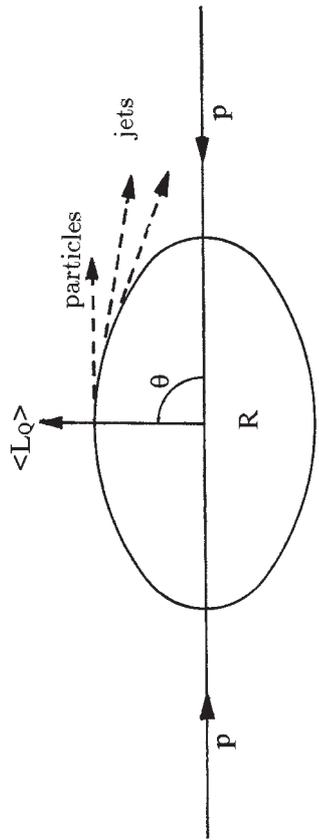
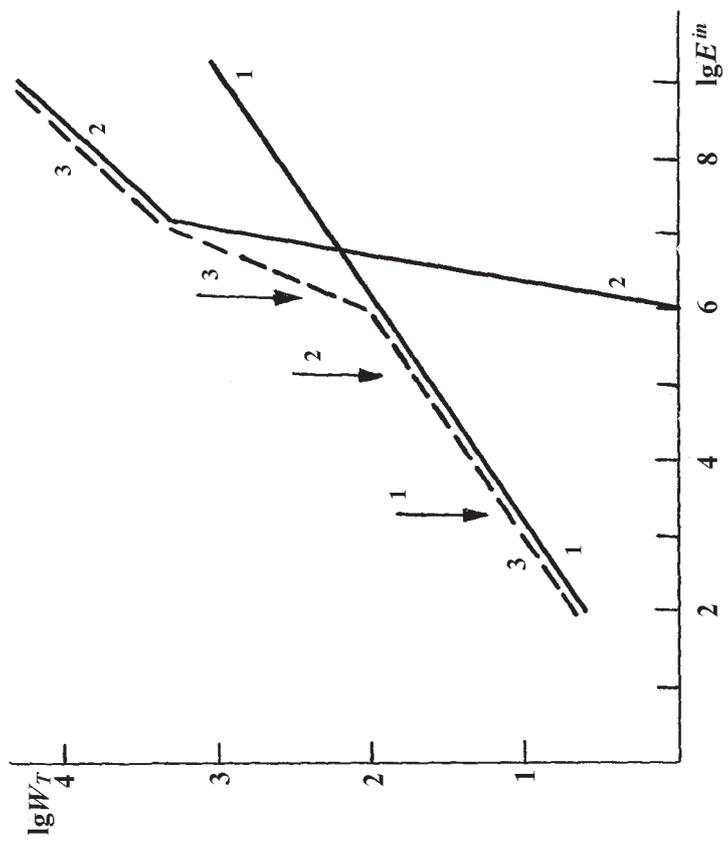
\end{document}